\documentclass[a4paper,twocolumn,11pt,accepted=2024-02-01]{quantumarticle}
\pdfoutput=1
\usepackage[utf8]{inputenc}
\usepackage[english]{babel}
\usepackage[T1]{fontenc}
\usepackage{amsmath}
\usepackage{amssymb}

\usepackage{verbatim}
\usepackage{rotating, booktabs}  
\usepackage{braket}
\usepackage{graphics}
\usepackage{graphicx}
\usepackage{hyperref}
\usepackage{color, xcolor}
\usepackage{upgreek}  
\usepackage{dcolumn}
\newcolumntype{z}[1]{D{.}{.}{#1}}
\usepackage{subfigure}
\usepackage{float}    

 \usepackage{microtype}
\usepackage{etoolbox}
 
\usepackage{comment}

 \usepackage{threeparttable}
 \usepackage{multirow}
 \usepackage{float}
 \usepackage{makecell}
 \usepackage{booktabs, tabularx, colortbl}
 \usepackage{array}
 \usepackage{enumerate}
\usepackage[linesnumbered,ruled ]{algorithm2e}
\SetKwInOut{Input}{Input}
\SetKwInput{Output}{Output} 

\usepackage{xargs}
\usepackage[colorinlistoftodos,prependcaption]{todonotes}

\newtheorem{definitionenv}{Definition}
\newtheorem{lemmaenv}[definitionenv]{Lemma}
\newtheorem{theoremenv}[definitionenv]{Theorem}
\newtheorem{corollaryenv}[definitionenv]{Corollary}
\newtheorem{propositionenv}[definitionenv]{Proposition}
\newtheorem{conjectureenv}[definitionenv]{Conjecture}
\newtheorem{remarkenv}[definitionenv]{Remark}
\newenvironment{remark}{\begin{remarkenv}\rm}{\end{remarkenv}}
\newcommand{\br}{\begin{remark}}
	\newcommand{\er}{\end{remark}}

\newtheorem{exampleenv}{Example}
\newtheorem{app-lemmaenv}[section]{Lemma}

\newenvironment{definition}{\begin{definitionenv}\rm}{\end{definitionenv}}
\newenvironment{lemma}{\begin{lemmaenv}\rm}{\end{lemmaenv}}
\newenvironment{theorem}{\begin{theoremenv}\rm}{\end{theoremenv}}
\newenvironment{corollary}{\begin{corollaryenv}\rm}{\end{corollaryenv}}
\newenvironment{example}{\begin{exampleenv}\rm}{\end{exampleenv}}
\newenvironment{proposition}{\begin{propositionenv}\rm}{\end{propositionenv}}
\newenvironment{conjecture}{\begin{conjectureenv}\rm}{\end{conjectureenv}}
\newenvironment{app-lemma}{\begin{app-lemmaenv}\rm}{\end{app-lemmaenv}}

\newcommand{\bd}{\begin{definition}}
	\newcommand{\ed}{\end{definition}}
\newcommand{\bl}{\begin{lemma}}
	\newcommand{\el}{\end{lemma}}
\newcommand{\elp}{\hspace*{\fill} $\Box$
\end{lemma}}
\newcommand{\bt}{\begin{theorem}}
\newcommand{\et}{\end{theorem}}
\newcommand{\etp}{\hspace*{\fill} $\Box$
\end{theorem}}
\newcommand{\bc}{\begin{corollary}}
\newcommand{\ec}{\end{corollary}}
\newcommand{\ecp}{\hspace*{\fill} $\Box$
\end{corollary}}
\newcommand{\bcj}{\begin{conjecture}}
\newcommand{\ecj}{\end{conjecture}}

\newcommand{\be}{\begin{example}}
\newcommand{\ee}{\end{example}}
\newcommand{\eep}{\hspace*{\fill} $\Box$
\end{example}}
\newcommand{\bp}{\begin{proposition}}
\newcommand{\ep}{\end{proposition}}
\newcommand{\epp}{
\end{proposition}}

\newcommand{\cB}{\mathcal{B}}

\newcommand{\cG}{\mathcal{G}}

\newcommand{\cN}{\mathcal{N}}

\newcommand{\cO}{\mathcal{O}}

\newcommand{\cR}{\mathcal{R}}
\newcommand{\cS}{\mathcal{S}}
\newcommand{\cT}{\mathcal{T}}

\newcommand{\tr}{\mathrm{tr}}

\newcommand{\cs}{C(\cS)}

\newcommand{\nsp}{\cN(\cS')}

\newcommand{\mo}{300n}
\newcommand{\moc}{\frac{1}{\epsilon^2} (n^2+2n)}
\newcommand{\mot}{n^4 2^t+ \frac{1}{\epsilon^2} n^4 }

\newcommand{\nc}{\newcommand}
\nc{\nnl}{\nn \\ &}  
\nc{\fot}{\frac{1}{2}} 
\nc{\oo}[1]{\frac{1}{#1}} 
\newcommand{\ben}{\begin{enumerate}}
\newcommand{\een}{\end{enumerate}}
\nc{\mc}{\mathcal}
\nc{\norm}[1]{\L\| #1 \R\|}

\nc{\onenorm}[1]{\L\| #1 \R\|_1} 
\nc{\vev}[1]{\langle#1\rangle}

\nc{\Ra}{\Rightarrow}
\nc{\zo}{\{0,1\}}	
\nc{\Adv}{\text{Adv}}
\nc{\eps}{\epsilon}

\newcommand{\kpsi}{\ket{\psi}}

\nc{\poly}{\mathsf{poly}}

\begin{document}

\title{Efficient learning of $t$-doped stabilizer states with single-copy measurements} 

\author{Nai-Hui Chia}
\affiliation{Department of Computer Science, Rice University, TX 77005-1892, United States}
\orcid{0000-0002-4138-7956}
 \author{Ching-Yi Lai}%
  \email{cylai@nycu.edu.tw}
 \affiliation{Institute of Communications Engineering, National Yang Ming Chiao Tung University,  Hsinchu 300093, Taiwan}
 \orcid{0000-0003-1970-8167}
\author{Han-Hsuan Lin}
 \affiliation{Department of Computer Science, National Tsing Hua University,  Hsinchu 30013, Taiwan}
\orcid{0000-0002-5126-0174}
\maketitle

\begin{abstract}
One of the primary objectives in the field of quantum state learning is to develop algorithms that are time-efficient for learning  states generated from quantum circuits. Earlier investigations have demonstrated time-efficient algorithms for states generated from Clifford circuits with at most $\log(n)$ non-Clifford gates. However, these algorithms necessitate multi-copy measurements, posing implementation challenges in the near term due to the requisite quantum memory. On the contrary, using solely single-qubit  measurements in the computational basis is insufficient in learning even the output distribution of a Clifford circuit with one additional $T$ gate under reasonable post-quantum cryptographic assumptions. In this work, we introduce an efficient quantum algorithm that  employs only single-copy measurements to learn states produced by Clifford circuits with a maximum of $O(\log(n))$ non-Clifford gates, filling a gap between the previous positive and negative results.
\end{abstract}

 \section{Introduction}

Quantum state tomography stands as a pivotal task in the realm of quantum information sciences \cite{Hra97, DPS03, BCG13}. In scenarios where numerous independent instances of an unknown state $\rho$ are available, the objective of a state tomography algorithm is to provide an output state $\hat{\rho}$ that closely approximates $\rho$. 
Deducing the requisite quantity of state copies and time resources to address this problem stands as a fundamental investigation in quantum information theory. 
 
It has been shown by Haah et al.~\cite{HWJ+17} and O'Donnell and Wright~\cite{OW16}  that the optimal sample complexity for the tomography of a quantum pure state  of dimension $d$  is $O(d/\epsilon^2)$, where $\epsilon$ represents the trace distance between $\hat{\rho}$ and $\rho$. These results indicate that the endeavor to learn an arbitrary $n$-qubit state mandates an exponential abundance of state copies
and an inherent investment of exponential time resources.

A natural question arises: which specific subsets of quantum states of $n$ qubits can be efficiently learned using at most $\mathsf{poly}(n)$ copies or even in $\mathsf{poly}(n)$ time? Intuitively, these subsets of quantum states should have concise representations. 
Chung and Lin~\cite{chung2021sample} showed that the set of states generated by  quantum circuits of polynomial size can be learned with $\mathsf{poly}(n)$ copies. This result indicates the feasibility of sample-efficient learning for a general class of states. 

Aaronson and Gottesman~\cite{AG04,AG08} and Montanaro~\cite{Mon17} considered the class of stabilizer states, where $n$-qubit states can be described in the stabilizer formalism using $O(n^2)$ bits~\cite{Got97}. 
Stabilizer states can be generated  by  Clifford circuits, comprising Hadamard, Phase, and controlled-NOT gates, on input computational basis states.
They showed that the sample complexity for learning stabilizer states is $O(n)$ and the time complexity is $O(n^3)$, given the capability to concurrently measure  multiple copies of a state.

 Building upon this foundation, a reasonable progression is to consider quantum states generated by Clifford circuits augmented with a few non-Clifford gates, such as the $\pi/8$-gate ($T$ gate). 
Notably, the Clifford gates and the $T$ gate constitute a universal gate set enabling universal quantum computation \cite{BMP+00}. 
 Lai and Cheng \cite{lai2022learning} initiated this exploration, demonstrating time-efficient learning algorithms for Clifford circuits with one layer of $\log(n)$ $T$ gates~\cite{lai2022learning}. 
 Arunachalam et al. introduced time-efficient algorithms for learning a category of constant-degree quantum phase states that also extend beyond the realm of stabilizer states~\cite{ABDY23}.

A quantum circuit is called a \textit{$t$-doped stabilizer circuit} if it comprises of Clifford gates and, at most, $t$ single-qubit non-Clifford gates. The resulting state of the circuit when applied to a computational basis state is referred to as a \textit{$t$-doped stabilizer state}.
 Following \cite{lai2022learning}, 
 it is shown that  {$t$-doped} stabilizer states with $t=O(\log(n))$   can be learned in polynomial time with polynomial samples by multiple concurrent results~\cite{GIKL23b,LOH23,HG23}.

Nonetheless, the algorithms  in~\cite{lai2022learning,GIKL23b,LOH23,HG23} require measurements on multiple copies of the states. This could potentially demand extra quantum resources during algorithm implementation, such as quantum memory.
On the contrary, Hinsche et al.~\cite{hinsche2022single} demonstrated that if restricted to single-qubit measurements in the computational basis, learning the output distribution of a 1-doped stabilizer circuit is hard under the assumption that quantum algorithms cannot efficiently solve the learning parity with noise  problem.

 Noticing the gap between the above results, we aim to investigate the following question: Using only single-copy measurements, whether there exist efficient  ($O(\poly(n,\epsilon))$) algorithms that use copies of an $n$-qubit $t$-doped stabilizer state $\ket{\psi}$ with  $t=O(\log(n))$ and generate a representation of a state that is  $\epsilon$-close to $\ket{\psi}$ in trace distance. We give an affirmative answer to this question in this paper and our main theorem (Theorem~\ref{thm:final-learn}) provides an efficient algorithm relying solely on single-copy measurements.

A key observation from \cite{lai2022learning,GIKL23b,LOH23,HG23} is that an $n$-qubit {$t$-doped}  stabilizer state is encompassed within a stabilizer code space of dimension at most $2^{2t}$, so we can efficiently map it to a $2t$-qubit state space (refer to Lemma~\ref{cor:decoding} and Lemma~\ref{lem:doped}). Consequently, the task of learning a {$t$-doped} $n$-qubit stabilizer state reduces to the identification of its corresponding stabilizer group and the related $2t$-qubit state. When $t=O(\log(n))$, the state can then be efficiently learned using a state tomography algorithm. To identify the corresponding stabilizer group, Bell sampling on multiple copies of the target state was used in \cite{lai2022learning,GIKL23b,LOH23,HG23}, but we seek algorithms that exclusively rely on single-copy measurements. Fortunately, Aaronson and  Gottesman~\cite{AG08} have demonstrated that learning an $n$-qubit stabilizer state can be achieved using just single-copy measurements, requiring mere $O(n^2)$ copies of the target state. Our algorithm adapts this method to pinpoint the associated stabilizer group of a $t$-doped stabilizer state.

\section{Stabilizer Formalism}\label{sec:intr_flag}
We introduce the basics of stabilizer formalism along with the corresponding notation.

 A single-qubit state space has a (computational) basis   \{$|0\rangle$, $|1\rangle$\}. 
The Pauli matrices  in the computational basis are  $I=\left(\begin{array}{cc}
  	1 & 0\\	0 & 1  
  \end{array}\right)$, $X=\left(\begin{array}{cc}
  0 & 1\\
  1 & 0
\end{array}\right)$,  $Z=\left(\begin{array}{cc}
  	1 & 0\\
  	0 & {-1}
  \end{array}\right)$, and $Y=iXZ$. 
Let $\cG_n=\Big\{ c M_1\otimes\cdots\otimes M_n:$  $M_i\in\{I$, $X,$ $Y,$ $Z\},$ $c\in\{\pm 1, \pm i\} \Big\}$ denote the $n$-fold Pauli group,
which has $2n$ independent Pauli generators up to a phase.
Any two Pauli operators either commute or anticommute with each other.

An $n$-qubit state $\ket{\psi}$ is said to be stabilized by a Pauli operator $g\in\cG_n$
if $g\ket{\psi}=\ket{\psi}$.
A stabilizer group  is an Abelian subgroup of $\cG_n$ that does not contain the minus identity.
The \textit{rank} of a stabilizer group is referred to as the smallest number of independent generators that span the stabilizer group.

Suppose that   $\cS$ is a stabilizer group of rank $r$  for some $r\leq n$ and $|\cS|=2^r$.
Let $\cs$ denote the subspace
\begin{align}
    \cs=&\Big\{\ket{\psi}\in\mathbb{C}^{2^n}: g\ket{\psi}=  \ket{\psi},\ \forall g\in\cS
\Big\}.
\end{align}
It can be shown that  $\cs$ is a complex vector space of dimension $2^{n-r}$ and
we can view $\cs$ as the code space of a quantum code which encodes  $n-r$ logical qubits into $n$ physical qubits~\cite{Got97}.

There exists a one-to-one correspondence between the code space $\cs$ and the raw $(n-r)$-qubit space
through the encoding and decoding Clifford circuits.
Moreover, the process of finding the encoding Clifford circuit, which consists of Hadamard, phase, and controlled-NOT gates, for a given stabilizer group is well-established~\cite{CG96,NC00}. 

\bl \label{cor:decoding}
For any $n$-qubit state $\ket{\psi}\in\cs$ with $|\cS|=2^r$,  we have  
\begin{align}
\ket{\psi}= V\Big(\ket{0}^{\otimes r}\otimes \ket{\phi}\Big),
\end{align}
where $\ket{\phi}$ is an $(n-r)$-qubit state and $V$ is a Clifford circuit.
Furthermore,   $V$ can be efficiently constructed in $O(rn^2)$ time if a set of generators for $S$ are given. 
\el

If $\cS$ is of rank $n$,
the subspace $C(\cS)$ is of dimension one and is called  a \textit{stabilizer state}.

\bd
For an $n$-qubit state $\ket{\psi}$,  the maximum integer $r$ such that $\ket{\psi}\in \cs$ for some stabilizer group $\cS\subset\cG_n$ of rank $r$ 
is called the \textit{stabilizer dimension} of $\kpsi$. 
\ed

It can be shown  that in a stabilizer group, at most two independent stabilizer generators do not commute with a specific single-qubit non-Clifford gate. Consequently, the following result can be established.

\bl \cite[Lemma 4.2]{grewal2023improved} \label{lem:doped}
Every $n$-qubit $t$-doped stabilizer state has stabilizer dimension at least $n-2t$.
\el

We are interested in learning quantum states of high stabilizer dimensions as   they exhibit behaviors closely resembling stabilizer states.

We define the distance between a quantum state $\ket{\psi}$ to a stabilizer group $\cS$ as 
\begin{align}
  D(\kpsi,\cS ) \triangleq \inf_{\ket{\phi}\in\cs} D(\kpsi,\ket{\phi}),
\end{align}
 where $D(\rho,\sigma)$ denotes the trace distance between the density operators of two quantum states $\rho$ and $\sigma$~\cite{NC00}:
\begin{align}
D(\rho,\sigma)= \max_{0\leq P\leq I} \tr(P(\rho-\sigma)).  \label{eq:trace_distance}  
\end{align}

Our approach relies solely on projective measurements of Pauli operators with eigenvalues $\pm 1$. Let $\ket{\psi}\in\cs$, where $\mathcal{S}\subset \mathcal{G}_n$ is a stabilizer group.
Consider a Pauli operator $g\in\mathcal{G}_n$ with eigenvalues $\pm 1$. The projectors onto the eigenspaces of $g$ are given by $\frac{1}{2}(I\pm g)$. If $\ket{\psi}$ is stabilized by $g$ (or $-g$), measuring $g$ on $\ket{\psi}$ will yield outcome $+1$ (or $-1$) with certainty. However, if $g$ anticommutes with one of the stabilizers of $\ket{\psi}$, measuring $g$ on $\ket{\psi}$ will result in outcome $+1$ with a probability of 1/2 and outcome~$-1$ with a probability of 1/2.

In the following discussion, instead of $\pm 1$, measurement outcomes will be denoted by binary values $0$ or~$1$, corresponding to $(-1)^0$ or~$(-1)^1$, respectively, for the ease of notation.

\bl(Gentle measurement~\cite{Win99}) \label{lemma:gm}
Consider a quantum state  $\ket{\psi}$  and a positive operator $M$ with eigenvalues no larger than one.
Suppose  that $\bra{\psi}M\ket{\psi}\geq 1-\epsilon$,
where  $0\leq \epsilon\leq 1$. Then the post-measurement state is $2\sqrt{\epsilon}$-close to $\ket{\psi}$  in trace distance.  
\el

\bl  \label{lemma:comm}
Consider two Pauli operators $g,h\in\cG_n$. Suppose that 
$\tr\left(\frac{I+g}{2} \ket{\psi}\bra{\psi}\right) \geq 0.99$ and 
    $\tr\left(\frac{I+h}{2} \ket{\psi}\bra{\psi}\right) \geq 0.99$
for some state $\ket{\psi}$. Then $g$ and $h$ commute.
\el
 \noindent\textbf{Proof.}  
 Let $\rho=\ket{\psi}\bra{\psi}$.  Let $\sigma$ and $\tau$ be the states resulting from measuring $\frac{I+g}{2}$ and $\frac{I+h}{2}$ on $\rho$, respectively. 
Clearly, we have $g\sigma g^\dag= \sigma $ and $h\tau h^\dag=\tau$.

Based on the assumption of the statement and the gentle measurement lemma (Lemma~\ref{lemma:gm}), we know that $D(\rho,\sigma)\leq 0.2$ and $D(\rho,\tau)\leq 0.2$. As a result, $D(\sigma,\tau)\leq 0.4$ by the triangle inequality.
Consequently,  we have
$
\tr\left( \frac{I+g}{2}\tau \right)\geq \tr\left( \frac{I+g}{2}\sigma \right)-0.4\geq 0.59. 
$

However, if $g$ and $h$ anticommute, we would have $\mathrm{tr}\left( \frac{I+g}{2}\tau \right)=\mathrm{tr}\left( \frac{I+g}{2}h \tau  h^\dag\right)=\mathrm{tr}\left( h\frac{I-g}{2} \tau  h^\dag \right)=\mathrm{tr}\left( \frac{I-g}{2}\tau \right).$ Thus $\mathrm{tr}\left( \frac{I+g}{2}\tau \right)=0.5<0.59$, which leads to a contradiction. Therefore, $g$ and $h$ commute.    \hfill     $\square$\\

\bl  \label{lemma:dev}
Let $\cS\subset\cG_n$ be a stabilizer group of $r$ independent generators $g_1,\dots,g_r$.
Suppose that $D(\ket{\psi},\cS)>\epsilon$. 
Then measuring $g_1,\dots, g_r$ on $\ket{\psi}$ returns outcome $0^r$ with probability at most
\[
1-\epsilon^2.
\]
\el
 \noindent\textbf{Proof.}  
Assume that 
\[
\ket{\psi}=\alpha \ket{\phi}+\beta \ket{\phi^\perp},
\]
where $\ket{\phi}\in C(\cS)$,  and $\vev{\phi|\phi^\perp}=0$.
Since $D(\ket{\psi},\cS)>\epsilon$, we  have $|\beta|>\epsilon$ and $|\alpha|^2< 1-\epsilon^2$. 
 
Let $P$ denote the projector onto $C(\cS)$. Then the probability of outcome $0^r$ by measuring $g_1,\dots,g_r$ on $\ket{\psi}$ is $\bra{\psi}P\ket{\psi}=|\alpha|^2< 1-\epsilon^2.$

    \hfill $\square$\\

\section{Learning quantum states of high stabilizer dimension}
Herein we provide an algorithm to approximately identify the stabilizer group $\cS$ 
from provided copies of a state $\ket{\psi}\in\cs$.

For two stabilizer groups $\cS$ and $\cT$, we denote 
\begin{align}
\cS\cap \cT\triangleq \{ g\in\cG_n: g\in\cS, g \text{ or }-g \in \cT   \}.
\end{align}

    \bl \label{lemma:dn}
    Let $r<n$, given  a stabilizer group $\cS\subset \cG_n$  of size $2^r$,
choose   a  stabilizer group $\cT\subset\cG_n$  of size $2^n$ at random.  Then 
\begin{align}
    \mathrm{Pr}\big\{ \cS\cap   \cT  \neq \{I\} \big\} \geq  \frac{1}{2^{n-r+1}+1}.
\end{align}
\el
\noindent\textbf{Proof.}  First, we consider the case where a stabilizer group $\mathcal{S}'\subset \mathcal{G}_n$ of rank $n$ is given, and a stabilizer group $\mathcal{T}'\subset\mathcal{G}_n$ of rank $r$ is chosen randomly.

We start by counting the number of stabilizer groups of rank $r$ in the following manner. To choose the first generator, we have $2^{2n}-1$ options. Then we choose the second generator from an equivalent space of $2^{2n-1}$ operators, half of which anticommute with the first generator. 
Thus the number of nontrivial options for the second generator is $2^{2n-2}-1$. 
Consequently, the number of distinct stabilizer groups in $\mathcal{G}_n$ of rank $r$ is given by 
 $${\left(2^{2n}-1\right)\left(2^{2n-2}-1\right)\cdots \left(2^{2n-2(r-1)}-1\right)}.$$

Next, we want to determine the number of stabilizer groups of rank $r$ that have only trivial intersections with the given stabilizer group $\mathcal{S}'$. To choose the first generator, we have $2^{2n}-2^n$ options since $|\mathcal{S}'|=2^n$. Then we choose the second generator from an equivalent space of $2^{2n-1}$ operators, half of which anticommute with the first generator. Note that half of the operators in $\mathcal{S}'$ also anticommute with the first generator. 
Thus the nontrivial options for the second generator is $2^{2n-2}-2^{n-1}$. 
Consequently, the number of stabilizer groups of rank $r$ that have only trivial intersections with $\mathcal{S}'$ is given by:
\begin{align*} 
&\prod_{j=0}^{r-1} \left(2^{2n-2j}-2^{n-j}\right). 
\end{align*}
Therefore, the probability that a randomly chosen $\mathcal{T}'$ and the given $\mathcal{S}'$ have a nontrivial intersection is:


 \begin{align*}
    \mathrm{Pr}&\big\{ \cS'\cap \cT' \neq \{I\} \big\} 
      =   1- \frac{\prod_{j=0}^{r-1} \left(2^{2n-2j}-2^{n-j}\right)}{\prod_{j=0}^{r-1} \left(2^{2n-2j}-1\right)}\\
   &\geq  1- \frac{ 2^{2n-2(r-1)}-2^{n-(r-1)}}{2^{2n-2(r-1)}-1} =\frac{1}{2^{n-r+1}+1}. 
\end{align*}

 In the above proof, we use the fact that each nontrivial Pauli operator anticommutes with half of the $n$-fold Pauli operators.

    Given that any two stabilizer groups of the same size can be transformed into each other by a Clifford unitary, and Clifford unitaries preserve the commutation relations of Pauli operators, we can deduce that for any stabilizer groups $\mathcal{S}_1$ and $\cS_2$ both of rank ${r_1}$, the  number of stabilizer groups of rank ${r_2}$ that share only a trivial intersection with $\mathcal{S}_1$   
    is equal to the number of stabilizer groups of rank ${r_2}$ that share only a trivial intersection with $\mathcal{S}_2$, 
    for any values of $r_1$ and $r_2$.
By applying Bayes's theorem, we can establish the desired statement.
    
    \hfill $\square$\\

Next we show how to find the stabilizer group for  given copies of a quantum state through single-copy measurements on $O(n^2)$ copies  of the target state.  We extend the algorithm for learning stabilizer states presented in~\cite{AG08} to our specific context, outlined in Algorithm~\ref{Algorithm:I}.

 \begin{theorem}\label{thm:learn-cs}
 There is a quantum algorithm  that 
 identifies the stabilizer group associated with  an unknown $n$-qubit quantum state $\ket{\psi}$  of stabilizer dimension $r=n-t$, where $t=O(\log(n))$,
 given access to copies of $\ket{\psi}$. 
 For any $\epsilon>0$, the algorithm produces  a stabilizer group $\cS\subset \cG_n$ of rank at least $r$ such that
 $
 D(\ket{\psi},\cS)<\epsilon.
 $
The algorithm employs  single-copy measurements on ${O}(n^2 2^t +\frac{1}{\epsilon^2} n^2) $
 copies of $\ket{\psi}$, runs  in time  $O(\mot)$, and  fails with probability exponentially small in $n$.

 \end{theorem}

\noindent\textbf{Proof.}  
 Suppose that $\ket{\psi}\in C(\mathcal{S'})$.
To find $r=n-t$ independent generators for $\mathcal{S'}$, we can adapt the approach of sampling random stabilizer groups and determining their nontrivial intersections with the target stabilizer group $\mathcal{S'}$, as described in~\cite{AG08}. 

However, the situation is more complex in our case. 
Let $\nsp\triangleq\{ h\in\cG_n: hg=gh, \forall g\in\mathcal{S'}  \}$ denote the normalizer group  of $\mathcal{S'}$ in $\mathcal{G}_n$.
Since $\mathcal{S'}$ is of rank $r<n$,  $\nsp$ contains generators not in $\mathcal{S'}$ that commute with any operators in $\mathcal{S'}$. This additional complexity may affect the correctness of our algorithm.

For $\mathcal{S'}$ of rank $r=n-t$, according to Lemma~\ref{lemma:dn}, a random stabilizer $\mathcal{T}$ of size $2^n$ would have a nontrivial intersection with $\mathcal{S'}$ with probability $\frac{1}{2^{t+1}+1}$ . To achieve a failure probability of $2^{-n}$ for $O(n)$ randomly picked stabilizer groups to have nontrivial intersections with $\mathcal{S'}$, we need to make $O(n 2^t)$ random picks of $\mathcal{T}$ using the Hoeffding inequality. As $\mathcal{S'}$ has fewer than $n$ independent generators, its basis can be found from $O(n)$ elements in $\mathcal{S'}$ with a failure  probability of $2^{-n}$.

Next, we outline the procedure for finding the nontrivial intersection of $\mathcal{S'}$ and an arbitrary stabilizer group $\mathcal{T}$ with $n$ independent generators ${g_1, g_2,\dots, g_n}\subset \mathcal{G}_n$.

An arbitrary element in $\cT$ can be expressed as 
\begin{align}
    g(a)=&\prod_i  g_i^{a_i},
\end{align}
where   $a_i \in \{0,1\}$.    
Since $g_1, g_2,\dots, g_n$ commute with each other, we can sequentially measure them on one copy of $\ket{\psi}$. 
Let the measurement outcomes be $b_1, b_2,\dots, b_n\in \{0,1\}$, respectively. 
Then, we can compute the binary outcome of measuring $g(a)$ on $\ket{\psi}$ as:
\begin{align}
    b\cdot a=\sum_i a_i b_i,
\end{align}
which can be computed for any $a \in \{0,1\}^n$ from the single-copy measurement outcomes $b_1, b_2,\dots, b_n$.

The following cases describe the possible values of $b\cdot a$ based on the properties of $g(a)$ and its relation to stabilizers of $\ket{\psi}$:
 a) $b\cdot a=0$ (respectively, $b\cdot a=1$) with certainty if $\ket{\psi}$ is stabilized by $g(a)$ (respectively, $-g(a)$). b) $b\cdot a=0$ or $1$ with probabilities 1/2 and 1/2, respectively, if $g(a)$ anticommutes with a stabilizer of $\ket{\psi}$.
 c) $b\cdot a=0$ or $1$ with nontrivial probabilities if $g(a)\in \nsp\setminus \pm \mathcal{S'}$.

Suppose that each of $g_1, g_2,\dots, g_n$ is measured on $m$ copies of $\kpsi$. 
Let $b_{i,j}\in\{0,1\}$ be the binary of measurement outcome  of $g_j$ on the $i$-th copy of $\kpsi$.
If $h(a)$ is a stabilizer of $\ket{\psi}$ for some $a\in\{0,1\}^n$, 
then  $a\in\{0,1\}^n$ is a solution to the following linear system of equations: 
\begin{align}
    B a  =0^{m},  \label{eq:ls}
\end{align}
where $B=[b_{i,j}]\in\{0,1\}^{m\times n}$. 
Note that  for the sake of analysis convenience, we do not consider solutions to $B a = 1^{m}$ such that $-h(a)$ is a stabilizer, but this exclusion does not affect the overall complexity, which remains unchanged up to a constant.

 Now consider a Pauli operator $h(a)\notin  \cS$  with $\bra{\psi}\frac{I+h(a)}{2}\ket{\psi}<0.99$.
If $Ba=0^m$, then $h(a)$ could be chosen as a solution to Step~2)-c), which can cause Algorithm~\ref{Algorithm:I} to fail.
Choosing $ m \geq 300n$, the probability of such an  event is  given by:
\begin{align*}
     &\textrm{Pr} \Big\{ Ba=0^m \Big| h(a)\notin  \cS \text{ and }  \bra{\psi}\frac{I+h(a)}{2}\ket{\psi}<0.99 \Big\}\\
     &\leq (0.99)^m \approx  e^{-0.01 m}=e^{-3n}.
\end{align*}
Since $|\mathcal{T}|=2^n$ and Step~2) is repeated $10n 2^t=\poly(n)$ times, by applying the union bound, the probability of Step~2) failing is at most $e^{-n}$.

By Lemma~\ref{lemma:comm} and the choice of $m$, the generators in $\mathcal{B}$ at Step~3) will commute with a probability of at least $1-e^{-n}$.

In Step 4), the basis $\mathcal{B}$ is checked by measuring $m'$ copies of $\ket{\psi}$ in the basis of $\mathcal{B}$, where $m'$ is to be determined. If any outcome of $1$ is observed, a Gaussian process can be performed to remove the independent generators that contribute to the outcomes of $1$s.

It remains to  bound the probability that the algorithm outputs a stabilizer group $\cS$ with $D(\ket{\psi}, \cS) > \epsilon$.
Such a stabilizer group would pass the verification step 4) if the outcomes of measuring $m'$ copies of $\ket{\psi}$ in the basis of $\mathcal{B}$ are all $0$s.
 According to Lemma~\ref{lemma:dev}, measuring a basis of $\mathcal{S}$ on $\ket{\psi}$ will return outcome $0^r$ with probability at most $1-\epsilon^2$.  
By choosing $m'\geq  \moc$,  the probability of this event is no larger than  $(1-\epsilon^2)^{ m'} \approx e^{-m' \epsilon^2} \leq    e^{-n^2-2n}$.
As there are no more than $2^{n^2+n}$ different stabilizer groups, the total probability of  these error events is no larger than $e^{-n}$ by the union bound.

As a result, the overall failure probability is $2^{-n}$.

Finally, we analyze the time complexity. 
In Step 2)-a), sampling a stabilizer group of rank $n$ can be done in $\cO(n^2)$ \cite{BM21,VDB21}.
In Step 2)-b), measuring an $n$-fold Pauli operator takes time $O(n)$. Thus, measuring $\ket{\psi}$ in the basis of a stabilizer group takes time $O(n^2)$, and each Step 2)-b) takes time $O(n^3)$.

Step 2)-c) also takes time $O(n^3)$ for Gaussian elimination. In Step 2)-d), determining whether $\mathcal{S}\cap \mathcal{T}\notin \langle \mathcal{R}\rangle$ can also be done by Gaussian elimination in $O(n^3)$.
Step 2) is repeated $O(n2^t)$ times, which takes time $O(n^42^t)$.

Step 3) takes time $O(n^3)$. 

Step 4) takes time $O(\frac{1}{\epsilon^2}n^4)$ as measuring each copy in the basis of $\cB$ takes time $O(n^2)$.
 
Therefore, the overall time complexity is $O(\mot  )$.

       \hfill $\square$\\

\begin{algorithm}[h] \small
	\setcounter{AlgoLine}{0}
	\Input{${O}(n^2 2^t +\frac{1}{\epsilon^2} n^2)$ copies of a state $\ket{\psi}$ of stabilizer dimension $n-t$ where $t=O(\log(n))$
 and a parameter $\epsilon$.}
	\Output{A set of independent generators for a stabilizer group $\cS$ with $D(\ket{\psi},\cS)<\epsilon$.}
	\begin{enumerate}[1)]
        \item  Initialize $\cR= \{I\}$.
        \item Repeat the following steps $10n 2^t$ times:
        \begin{enumerate}[a)]
        \item Pick a stabilizer group $\cT$ of rank $n$ at random.
        \item Measure $\mo$ copies of $\ket{\psi}$ in the basis of $\cT$.
        \item Find a generating set for $\cS\cap\cT$ by solving Eq.~(\ref{eq:ls}).
        \item Add any elements of the generating set that are  not in $\langle \cR \rangle$  to $\cR$.
        \end{enumerate}
 \item  Find a basis  $\cB$ for $\cR$ by Gaussian elimination. 
 \item  Measure  $\moc$ copies of $\ket{\psi}$ in the  basis of $\cB$. 
Keep the generators that give all-zero outcomes. Return them as generators of $\cS$.

	\end{enumerate}

	\caption{Learning  an unknown stabilizer group of a quantum state} \label{Algorithm:I}
\end{algorithm}

\section{Learning $t$-doped stabilizer states}

To characterize an $n$-qubit {$t$-doped}  stabilizer state with stabilizer dimension  at least $n-2t$,
 one can specify its stabilizer encoding circuit along with the corresponding raw $2t$-qubit state, as described in Lemma~\ref{cor:decoding}. 
Utilizing Theorem~\ref{thm:learn-cs}, one can learn the stabilizer encoding circuit using single-copy measurements on copies of the target state. 
  Subsequently, the decoding circuit can be applied to the target state, thus obtaining its raw $2t$-qubit state. 
   In this context, we adopt the following fast state tomography algorithm to learn this $2t$-qubit state.
  
\begin{lemma}\label{lem:tomography} (fast state tomography~\cite{franca_et_al:LIPIcs.TQC.2021.7})
    There exists a quantum algorithm that for any unknown $n$-qubit pure state $\kpsi$,   uses $O(2^n n \log(\frac{1}{\delta})\epsilon^{-4})$-copies of $\kpsi$ and generates   {a representation of} a state $\ket{\tilde{\psi}}$ such that $D(\kpsi,\ket{\tilde{\psi}}) \leq \epsilon$ with probability $1-\delta$ in  $O(2^{2n} n^3 \log(\frac{1}{\delta})\epsilon^{-5})$ time. The algorithm  requires only single-copy Clifford measurements and classical post processing.
\end{lemma}

\nc{\nocthmn}{O(n^2 2^{2t}+\frac{2^{4t} n^2}{\epsilon^8})}    
\nc{\tthmn}{O(n^4 2^{2t}+\frac{2^{4t} n^4}{\epsilon^8})}

Our main result is as follows.
\begin{theorem}~\label{thm:final-learn}
    There exists a quantum algorithm that identifies an unknown  $n$-qubit $t$-doped   stabilizer state $\ket{\psi}$, where $t=O(\log(n))$, given access to copies of $\ket{\psi}$. 
    For any $\epsilon>0$, the algorithm generates a representation of a state $\ket{\psi'}$  such that $D(\ket{\psi},\ket{\psi'})<\epsilon$.
     
    The algorithm employs single-copy measurements on $\nocthmn$ copies of  $\ket{\psi}$,
    runs in time $\tthmn$,
    and fails with probability exponentially small in $n$.\footnote{ If $t=\log(n)$, the sample and time complexities are $O(n^6/\epsilon^8)$ and $O(n^8/\epsilon^8)$, respectively. If $t=O(\log(n))$, sample and time complexities are both polynomial in $n$.}
    
 \end{theorem}

\noindent\textbf{Proof.} It is mentioned that a $t$-doped stabilizer state has stabilizer dimension at least $n-2t$.
 By Theorem~\ref{thm:learn-cs}, we can learn an approximate stabilizer group $\cS$ of $\kpsi$ such that $D(\ket{\psi},\cS)<\epsilon'=  O(\frac{\epsilon^4}{2^{2t}})$ with $\nocthmn$ samples and $\tthmn$ time. 
 
  Let the rank of $\cS$ be $r'$. 
By measuring  $  O(2^{2t} \epsilon^{-4})$ copies of $\ket{\psi}$ in the basis of $\cS$, we will expect that  due to the choice of $\epsilon'$, the outcomes will be  all 0s with a negligible probability of failure.  
  Let the post measurement state be $\ket{\psi'}$ and we have  $ O(2^{2t} \epsilon^{-4})$ copies of $\ket{\psi'}$ from the previous measurements. 
 
 By Lemma~\ref{cor:decoding}, we can find the encoding circuit $V$ for the stabilizer group $\cS$.
 Applying $V^\dag$ to each copy of $\ket{\psi'}$, we get copies of $\ket{0}^{r'}\otimes \ket{\phi}$, where $\ket{\phi}$ has $2t$-qubits and this process takes $ O(2^{2t} \epsilon^{-4}) \times O(n^3)$ time. 
 With $O(2^{2t} \epsilon^{-4})$ copies of $\ket{\phi}$, we can apply Lemma~\ref{lem:tomography} to get an estimate $\ket{\phi'}$ with $O(\epsilon)$ error. 
 
 Finally, our estimate  of $\kpsi$ is $V(\ket{0}^{r'}\otimes \ket{\phi'})$, which is $O(\epsilon)$-close  to $\ket{\psi'}$ in trace distance and thus $O(\epsilon)$-close  to $\kpsi$ in trace distance.
    \hfill $\square$\\

\section{Discussion}
The existence of an efficient learning algorithm for states generated from Clifford circuits equipped with a polynomial count of $T$ gates  would imply the feasibility of efficiently learning states generated by quantum circuits of polynomial size.
Our algorithm extended previous results to learning quantum states generated from Clifford circuits augmented with $\log(n)$ non-Clifford gates,
using only single-copy measurements.
This exploration holds the potential to usher in practical applications in quantum physics and quantum computing, such as comprehending noise within quantum devices and delving into the correlations present within enigmatic quantum systems.

A natural follow-up question is whether using arbitrary single-qubit measurements, efficiently learning $t$-doped stabilizer states is possible. This will fill the gap between our result and \cite{hinsche2022single}.

Our technique applies to quantum process tomography~\cite{MRL08} of $t$-doped stabilizer circuits as well.
More specifically, one can efficiently identify the Choi–Jamiołkowski state~\cite{CHOI1975285,JAMIOLKOWSKI1972275} of $t$-doped stabilizer  circuit using single-copy measurements with polynomially many uses of the circuit. 
However, note that efficiently deriving a circuit decomposition from a provided representation of a Choi–Jamiołkowski state is not a straightforward task, especially when dealing with a combination of Clifford and non-Clifford gate sets.

While preparing this manuscript, we became aware of a similar result   independently discovered by Grewal et al.~\cite{GIKL23c}. Our main results in Theorem~\ref{thm:learn-cs}  and Algorithm~\ref{Algorithm:I} correspond to  Theorem 5.9 and Algorithm~1 in \cite{GIKL23c}, respectively. However, our algorithm exhibits   improved time and sample complexities, particularly with a better dependence on the number of qubits $n$.  Learning a $t$-doped stabilizer state involves identifying its corresponding (non-full rank) stabilizer group. We demonstrate that this can be achieved through single-copy measurements, extending the algorithm introduced by Aaronson and Gottesman~\cite{AG08}. Lemma~\ref{lemma:dn} establishes that a random stabilizer group and the target stabilizer group have a nontrivial intersection with a nontrivial probability. Subsequently, in Theorem~\ref{thm:learn-cs}, single-copy measurement outcomes in the basis of a random stabilizer group can be utilized to deduce a stabilizer generator of the target stabilizer group with high probability. As a result, the time complexity dependence on $n$ in our algorithm aligns with that in~\cite{AG08}. On the contrary, Grewal et al.~\cite{GIKL23c} choose a different strategy. They introduce a computational difference sampling based on the outcomes of single-copy measurements to emulate the Bell difference sampling, a method they had previously demonstrated for learning the target stabilizer~\cite{GIKL23b}. This approach relies heavily on a complex boolean Fourier analysis.

\section*{Acknowledgements}
 NHC was supported by NSF award FET-2243659, Google Scholar Award, and DOE award DE-SC0024301. 
 
 CYL was supported by the National Science and Technology Council (NSTC) in
 Taiwan, under Grant 111-2628-E-A49-024-MY2, 112-2119-M-A49-007, and 112-2119-M-001-007.
 
 HHL was supported by NSTC QC project under Grant no.  111-2119-M-001-004- and   110-2222-E-007-002-MY3.

\bibliographystyle{quantum}
\bibliography{ref}

\end{document}